# Multi-Objective Digital PID Controller Design in Parameter Space

Haoan Wang, Levent Guvenc

*Abstract*—This paper presents a multi-objective digital PID controller design method using the parameter space approach of robust control. Absolute stability is treated first by finding the digital PID controller gain parameter space corresponding to closed loop poles being inside the unit circle. Phase margin, gain margin and a mixed sensitivity bound are treated as frequency domain constraints. Determination of digital PID controller parameter space regions satisfying these constraints is presented. All of these regions are superimposed to obtain a multi-objective digital PID controller gain parameter space solution region. The path following controller design of an automated driving vehicle is used as an example to illustrate the method.

## I. INTRODUCTION

The parameter space approach is a part of the parametric approach to robust control [1-3]. Using the method of mapping frequency domain bounds to the chosen parameter space, the method can be applied to treat frequency domain uncertainty as well [2-4]. The parameter space approach is computationally fast, has the advantage of obtaining solution regions rather than one set of controller gains and can easily handle time delays but lacks from the need to pre-specify the controller structure and being able to handle only two parameters at a time [3,5]. Parameter space robust control has recently been used successfully in a large number of applications ranging from yaw stability control and steering control to control of actuation in atomic force microscopy [6-10]. Several researchers have applied the parameter space approach to continuous time PID controllers for which the above-mentioned weaknesses of the parameter space approach are overcome as the controller structure is fixed and as there are only three controller parameters to be tuned, i.e. the proportional, integral and derivative gains [11-13].

Although there is previous work on continuous time PID controller design in parameter space, corresponding results are missing for digital PID controllers. While it is always possible to design a continuous time PID controller and then discretize it for a digital implementation, it is preferable to directly design the digital PID in the z-domain especially in the presence of a sampling time that is not too small which is typical for automotive control systems that rely on measurements from the CAN bus. This paper, therefore, focuses on a direct multi-objective digital PID design in the z-domain for absolute stability and for satisfying desired gain margin, phase margin and mixed sensitivity bound constraints.

The rest of paper is organized as follows. Section II presents the parameter space approach based robust PID controller design in the z-domain, where absolute stability, phase margin constraint, gain margin constraint and mixed sensitivity constraint are considered. In Section III, an example is used for illustrating multi-objective z-domain robust PD control satisfying phase margin and mixed sensitivity bound constraints simultaneously. The designed robust digital PD controller is applied to the autonomous vehicle path following control system in a simulation analysis. The paper ends with conclusions in Section IV.

## II. PARAMETER SPACE APPROACH IN Z-DOMAIN

### A. Absolute Stability in the z Domain

Let the characteristic equation of a feedback control system with control gains $k$ and uncertain parameters $q$ be given by $p(s, q, k) = 0$ in the $s$ domain. Hurwitz stability requires all roots of the characteristic equation $p(s, q, k) = 0$, to lie in the left-half plane. The parameter space solution is based on the Boundary Crossing Theorem which states that characteristic equation roots need to cross the stability boundary to go from stable to unstable ones and vice versa as parameters are changed [2]. The continuous time stability boundary can be crossed through the real root boundary (RRB), complex root boundary (CRB) or infinite root boundary (IRB) [2].

The corresponding absolute stability region in the z-domain is the inside of the unit circle. The Boundary Crossing Theorem is applicable again and transitions of characteristic equation roots from inside the unit circle (stable) to the outside of the unit circle (unstable) are only possible by crossing the unit circle as parameters are varied. In the z-domain there is only the complex root boundary CRB around the unit circle and the real root boundaries RRB at z=1 and z=-1. An infinite root boundary IRB does not exist for z-domain absolute stability.

The unit circle is given by

$$z = e^{sT} = e^{j\omega T} = e^{j\theta} = cos\theta + jsin\theta \quad (1)$$

H. Wang was with the Electrical and Computer Engineering Department, Ohio State University, Columbus, OH 43210 USA. (e-mail: wang.6184@osu.edu).

L. Guvenc is with the Automated Driving Lab and the Departments of Mechanical and Aerospace Engineering and Electrical and Computer Engineering, Ohio State University, Columbus, OH 43210 USA. (e-mail: guvenc.1@osu.edu).

where $\theta \equiv \omega T \in [0, 2\pi]$ and $T$ is the sampling time. Consider the standard digital PID controller given by

$$C(z) = k_p + k_i \frac{z}{z-1} + k_d \frac{z-1}{z} \quad (2)$$

where $k_p$, $k_i$ and $k_d$ are the proportional, integral and derivative gains, respectively. In the generic digital feedback control system shown in Fig. 1 with controller $C(z)$ and plant $G(z)$, the loop gain is the product of all transfer functions in the loop as

$$L(z) = C(z)G(z). \quad (3)$$

The closed loop transfer function is:

$$G_{cl}(z) = \frac{C(z)G(z)}{1 + C(z)G(z)} \quad (4)$$

The characteristic equation can be derived as

$$p(z, k, q) \equiv 1 + C(z)G(z) = 0 \quad (5)$$

Substitute (2) into (5) to obtain

$$(z-1)z + G(z)(k_p(z-1)z + k_i z^2 + k_d(z-1)^2) = 0 \quad (6)$$

as the characteristic equation of the digital PID controlled plant.

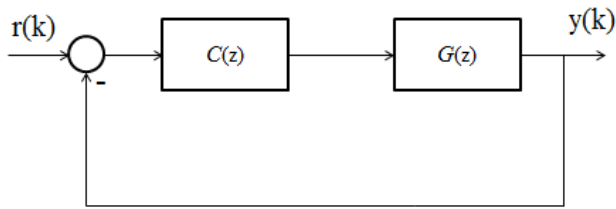

Fig. 1. Digital feedback control system block diagram

The complex root boundary CRB will be computed using Equation (6). Substitute the unit circle boundary $z = e^{j\theta} = \cos\theta + j\sin\theta$ and $G(z) = Re_G + jIm_G$ into (6) and separate the real part denoted by $Re$ and the imaginary part denoted by $Im$ of (6) to obtain the following two equations which can be used for calculating two free PID design parameters.

*Real part*:
$$(1 + (k_p + k_i + k_d)Re_G)\cos2\theta - (k_p + k_i + k_d)Im_G\sin2\theta - (1 + (k_p + 2k_d)Re_G)\cos\theta + (k_p + 2k_d)Im_G\sin\theta + Re_G k_d = 0 \quad (7)$$

*Imaginary part*:
$$(1 + (k_p + k_i + k_d)Re_G)\sin2\theta + (k_p + k_i + k_d)Im_G\cos2\theta - (1 + (k_p + 2k_d)Re_G)\sin\theta - (k_p + 2k_d)Im_G\cos\theta + Im_G k_d = 0 \quad (8)$$

A sweep of angle $\theta \in [0, 2\pi]$ is used to solve Equations (7) and (8) above for any two of the three digital PID controller gains. When one of the digital PID gains is zero in the case of PI or PD controllers, Equations (8) and (9) provide the solution region in the corresponding controller parameter space. When all three digital PID gains are present, it is possible to solve (7) and (8) for a grid of possible values of one of the control parameters and to obtain a three dimensional absolute stability solution region.

The real root boundary RRB is calculated using $z=1$ ($\theta=0°$) and $z=-1$ (($\theta=180°$) in Equations (7) and (8) above or Equation (6). Note that these equations will degenerate into a single equation for each of the two real root boundaries. For the real root boundary at $z=1$, the RRB equation is $k_i=0$. The RRB at $z=-1$ corresponds to a singular solution and the RRB equation is

$$2k_p + 4k_d + k_i = -\frac{1}{G(z)\big|_{z=-1}} \quad (9)$$

Equations (7) and (8) can be combined into the matrix equation

$$Ak \equiv \begin{bmatrix} a_{11} & a_{12} & a_{13} \\ a_{21} & a_{22} & a_{23} \end{bmatrix} \begin{bmatrix} k_p \\ k_i \\ k_d \end{bmatrix} = \begin{bmatrix} b_1 \\ b_2 \end{bmatrix} \equiv b \quad (10)$$

where

$a_{11} = Re_G \cos2\theta - Im_G(\sin2\theta + \cos\theta + \sin\theta)$
$a_{12} = Re_G(\cos2\theta + 1) - Im_G(\sin2\theta + 2\cos\theta + 2\sin\theta)$
$a_{13} = Re_G(\cos2\theta + 1) - Im_G(\sin2\theta + 2\cos\theta + 2\sin\theta)$
$a_{21} = Re_G(\sin2\theta + \sin\theta) + Im_G(\cos2\theta - \cos\theta)$
$a_{22} = Re_G(\sin2\theta + 2\sin\theta) + Im_G(\cos2\theta - 2\cos\theta + 1)$
$a_{23} = Re_G \sin2\theta + Im_G \cos2\theta$
$b_1 = -\cos(2\theta) + \cos(\theta)$
$b_2 = -\sin(2\theta) + \sin(\theta)$

Frequencies $\theta = \omega T$ that make rank($A$)=rank($[A:b]$)=1 (or $b \in$ range($A$)) are singular frequencies and result in infinitely many solutions corresponding to a line in the chosen space of two controller parameters.

As an example, consider the plant $G(z) = \frac{1}{z(z+1)}$ for PD and PI controller design. Fig. 2 and Fig. 4 illustrate 3D plots where $(k_p, k_i)$ and $(k_d, k_p)$ are scheduled by sampling time $T \in [0.3, 0.8]$ sec on the third vertical axis. Fig. 3 and Fig. 5 show detailed views of $k_p, k_i$ and $k_d, k_p$ parameter spaces and pole locations within the z-plane when sample time is 0.3 sec. The shaded blue area represents the stable region and it can be seen that when $(k_p, k_i)$ $(k_d, k_p)$ points are selected inside the stable region, on the stable boundary and outside the stable region, the corresponding two poles of the closed loop transfer function (4) are inside the unit circle, on the unit circle and outside the unit circle, respectively, as expected.

## B. Phase Margin Constraint in the z-Domain

From (1), we get the following equation:

$$z = e^{sT} = e^{j\omega T} = \cos\omega T + j\sin\omega T \quad (11)$$

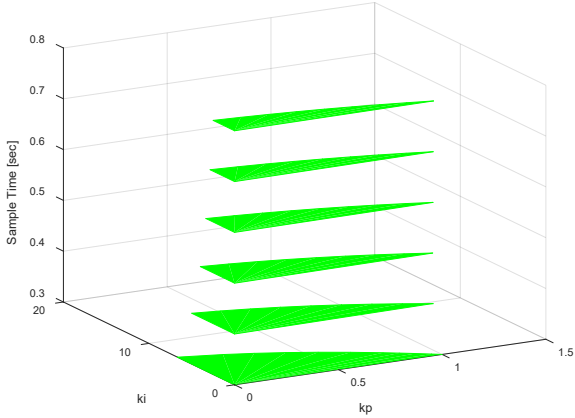

Fig. 2. Overall $k_p - k_i$ solution region scheduled by sample time $T$

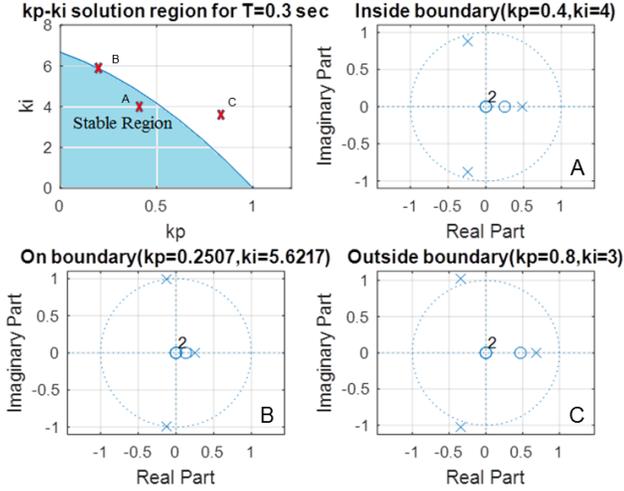

Fig. 3. $k_p, k_i$ solution region and poles location in z-plane

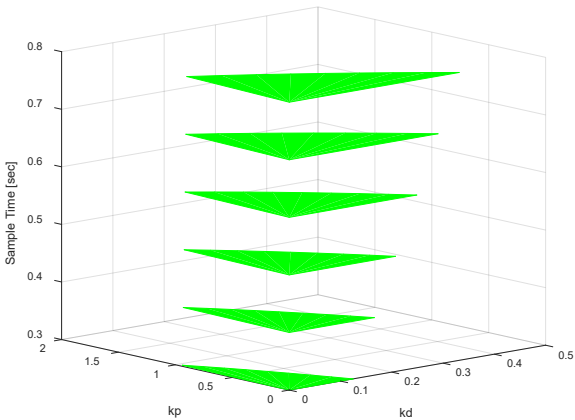

Fig. 4. Overall $k_d - k_p$ solution region scheduled by sample time $T$

Consider $\omega_{gc}$ as the gain crossover frequency where the loop gain $L(z)$ is unity or zero decibels as

$$|L(z)| = |L(e^{j\omega_{gc}T})| = 1 \quad (12)$$

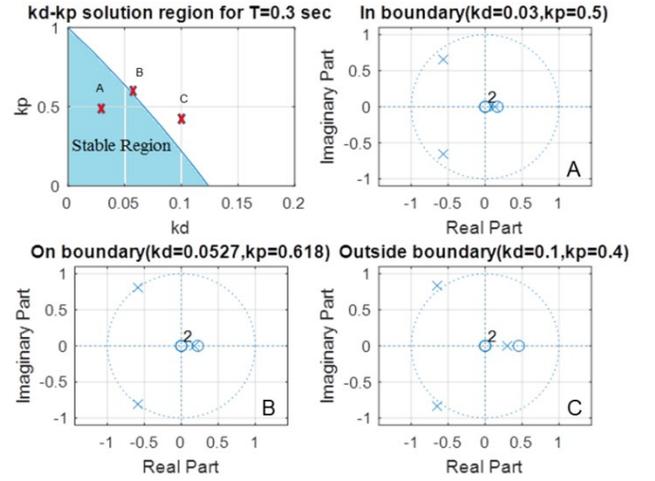

Fig. 5. $k_p, k_d$ solution region and poles location in z-plane

The expression of the phase margin $PM$ is

$$L(z) = e^{j(PM-\pi)} = -\cos(PM) - j\sin(PM) \quad (13)$$

Substituting from (3) into (13), the real and imaginary component equations of $L(z)$ are written as

$$Re(L(z)) = Re(C(z)G(z)) = -\cos(PM) \quad (14)$$

$$Im(L(z)) = Im(C(z)G(z)) = -\sin(PM) \quad (15)$$

Substituting (2) into (14), (15), we obtain

$$Re\left(\left(k_p + k_i\frac{z}{z-1} + k_d\frac{z-1}{z}\right)G(z)\right) = -\cos(PM) \quad (16)$$

$$Im\left(\left(k_p + k_i\frac{z}{z-1} + k_d\frac{z-1}{z}\right)G(z)\right) = -\sin(PM) \quad (17)$$

where the PID control gains in (16) and (17) can be expressed as follows:

$$\left.\left(k_p + k_i\frac{z}{z-1} + k_d\frac{z-1}{z}\right)\right|_{z=\cos\theta+j\sin\theta} =$$

$$= k_p + k_i\frac{(\cos\theta + j\sin\theta)}{(\cos\theta + j\sin\theta) - 1} + k_d\frac{(\cos\theta + j\sin\theta) - 1}{(\cos\theta + j\sin\theta)}$$

$$= k_p + k_i\frac{(\cos\theta + j\sin\theta)((\cos\theta - 1) - j\sin\theta)}{((\cos\theta - 1) + j\sin\theta)((\cos\theta - 1) - j\sin\theta)} +$$

$$\quad k_d\frac{(\cos\theta + j\sin\theta - 1)(\cos\theta - j\sin\theta)}{(\cos\theta + j\sin\theta)(\cos\theta - j\sin\theta)}$$

$$= k_p + k_i\frac{1 - \cos\theta - j\sin\theta}{(\cos\theta - 1)^2 + (\sin\theta)^2} + k_d(1 - \cos\theta + j\sin\theta) \quad (18)$$

Substituting (18) and $G(z) = Re_G + jIm_G$ into (16) and (17), Equations (19) and (20) are derived for solving the parameters of the PID controller which satisfy the phase margin constraint as

$$k_p Re_G + \frac{k_i(Re_G(1-\cos\theta)+Im_G\sin\theta)}{(\cos\theta-1)^2+(\sin\theta)^2} + k_d(Re_G(1-\cos\theta) - Im_G\sin\theta) = -\cos(PM) \quad (19)$$

$$k_p Im_G + \frac{k_i(Im_G(1-\cos\theta)-Re_G\sin\theta)}{(\cos\theta-1)^2+(\sin\theta)^2} + k_d(Re_G\sin\theta + Im_G(1-\cos\theta)) = -\sin(PM) \quad (20)$$

Using a grid of $\theta \in [0,2\pi]$, Equations (19) and (20) can be used to obtain the parameter pace region in any two of the digital PID gains when the other one is fixed.

## C. Gain Margin Constraint via z-Plane

Consider the phase crossover frequency $\omega_{pc}$ where the phase becomes $-180°$ given by

$$\angle L(z) = \angle L(e^{j\omega_{pc}T}) = -180° \quad (21)$$

$$L(z) = L(e^{j\omega_{pc}T}) = \frac{1}{M}\angle -180° = -\frac{1}{M} \quad (22)$$

where $M = 10^{(\frac{GM}{20})}$ and $GM$ is gain margin bound in decibels (dB). Substituting (3) into (22), the real and imaginary component Equations of $L(z)$ are written as

$$Re(L(z)) = Re(C(z)G(z)) = -\frac{1}{M} \quad (23)$$

$$Im(L(z)) = Im(C(z)G(z)) = 0 \quad (24)$$

Substituting (2) into (23) and (24), we obtain:

$$Re\left((k_p + k_i\frac{z}{z-1} + k_d\frac{z-1}{z})G(z)\right) = -\frac{1}{M} \quad (25)$$

$$Im\left((k_p + k_i\frac{z}{z-1} + k_d\frac{z-1}{z})G(z)\right) = 0 \quad (26)$$

Substituting (18) and $G(z) = Re_G + jIm_G$ into (25) and (26), Equations (27) and (28) are derived as

$$k_p Re_G + \frac{k_i(Re_G(1-\cos\theta)+Im_G\sin\theta)}{(\cos\theta-1)^2+(\sin\theta)^2} + k_d(Re_G(1-\cos\theta) - Im_G\sin\theta) = -\frac{1}{M} \quad (27)$$

$$k_p Im_G + \frac{k_i(Im_G(1-\cos\theta)-Re_G\sin\theta)}{(\cos\theta-1)^2+(\sin\theta)^2} + k_d(Re_G\sin\theta + Im_G(1-\cos\theta)) = 0 \quad (28)$$

and can be used for solving two parameters of the digital PID controller which satisfy the gain margin constraint.

## D. Mixed Sensitivity Constraint in the z-Domain

Mixed sensitivity design aims to map frequency domain criteria of robust control into parameter space, which must satisfy the following robust performance requirement

$$\||W_S S| + |W_T T|\|_\infty < 1 \text{ or } |W_S S| + |W_T T| < 1, \forall \omega \quad (29)$$

where $S = 1/(1+L)$ and $T = L/(1+L)$ are sensitivity and complementary sensitivity functions and $W_S$ and $W_T$ are corresponding weights. Different choices of weight functions $W_s(s)$ and $W_T(s)$ in the s domain were introduced in [3]. Similar first order discrete time weight transfer functions $W_s(z)$ and $W_T(z)$ can be derived using the zero-order hold method from these or they can be designed directly in discrete time.

The mixed sensitivity constraint can be expressed as

$$|W_s(z)| + |W_T(z)L(z)| = |1 + L(z)| \quad (30)$$

$$L(z) = |L(z)|\angle \theta_L = |L(z)|e^{j\theta_L} \quad (31)$$

where the solution of $|L(z)|$ can be expressed as:

$$|L(z)| = \frac{-\cos\theta_L + |W_s(z)||W_T(z)| \pm \sqrt{\Delta}}{1 - |W_T(z)|^2} \quad (32)$$

where

$$\Delta = \cos^2\theta_L + |W_s(z)|^2 + |W_T(z)|^2 - 2|W_s(z)||W_T(z)|\cos\theta_L - 1, \theta_L \in [0,2\pi], \Delta \geq 0 \quad (33)$$

$L(z)$ can be presented in terms of a controller K as shown in the equation:

$$L(z) = K(z)G(z) = (K_R + jK_I)G(z) \quad (34)$$

which can be used to solve for the real part $K_R$ and imaginary part $K_I$ of the controller. Based on the PID controller expression, Equation (35) is derived as:

$$K_R + jK_I = k_p + k_i\frac{z}{z-1} + k_d\frac{z-1}{z} \quad (35)$$

Separating the real and imaginary parts of Equation (34) and substituting into Equation (35), the following Equations (36) are derived

$$\begin{cases} K_R = k_p + k_i\frac{1-\cos\theta}{(\cos\theta-1)^2+(\sin\theta)^2} + k_d(1-\cos\theta) \\ K_I = -k_i\frac{\sin\theta}{(\cos\theta-1)^2+(\sin\theta)^2} + k_d\sin\theta \end{cases} \quad (36)$$

For PD controller parameters $k_d$ and $k_p$ Equations (36) become

$$k_d = \frac{K_I}{\sin\theta} \quad (37)$$

$$k_p = K_R - k_d(1-\cos\theta) = K_R - K_I\frac{(1-\cos\theta)}{\sin\theta} \quad (38)$$

For PI controller parameters $k_p$ and $k_i$, Equations (36) become

$$k_i = -\frac{K_I((\cos\theta-1)^2+(\sin\theta)^2)}{\sin\theta} \quad (39)$$

$$k_p = K_R - k_i \frac{1-\cos\theta}{(\cos\theta-1)^2+(\sin\theta)^2} = K_R + \frac{K_I(1-\cos\theta)}{\sin\theta} \quad (40)$$

For PID controller design, substituting from Equation (36) into (34), using $G(z) = Re_G + jIm_G$ and separating the real and imaginary parts, Equations (41) and (42) are derived

$$k_p Re_G + \frac{k_i(Re_G(1-\cos\theta)+Im_G\sin\theta)}{(\cos\theta-1)^2+(\sin\theta)^2} + k_d(Re_G(1-\cos\theta) - Im_G\sin\theta) = |L(z)|\cos\theta_L \quad (41)$$

$$k_p Im_G + \frac{k_i(Im_G(1-\cos\theta)-Re_G\sin\theta)}{(\cos\theta-1)^2+(\sin\theta)^2} + k_d(Re_G\sin\theta + Im_G(1-\cos\theta)) = |L(z)|\sin\theta_L \quad (42)$$

The equations presented in this section can be used to determine the PID controller gain parameter space regions where absolute stability, gain margin, phase margin and mixed sensitivity bounds are satisfied. This is illustrated with an example in the next section.

## III. MULTI-OBJECTIVE DIGITAL PD CONTROLLER DESIGN EXAMPLE

Consider the experimentally validated continuous time transfer function $G(s)$ of the Ford Fusion experimental vehicle model [14] of the Automated Driving Lab of the Ohio State University where $G(s)$ is from front wheel steering angle $\delta_f$ to lateral deviation $y$ form a desired path and is given in Equation (43) below. $G(z)$ is discretized from $G(s)$ using the zero order hold method with sampling time T=0.01 sec as given in Equation (44) below.

$$G(s) = \frac{227.6\,s^2 + 5536\,s + 36260}{s^4 + 22.16\,s^3 + 37.92\,s^2} \quad (43)$$

$$G(z) = \frac{0.01147\,z^3 - 0.008747\,z^2 - 0.01145\,z + 0.009058}{z^4 - 3.798\,z^3 + 5.397z^2 - 3.4\,z + 0.8012} \quad (44)$$

The phase margin constraint and mixed sensitivity constraint are taken into account simultaneously in the digital PD controller design. The phase margin is required to be within PM ∈ [20,80] deg and the parameters for mixed sensitivity constraint are: low frequency bound $l_s = 0.5$, the high frequency bound $h_s = 4$, and the approximate bandwidth was $\omega_s$=5rad/s for sensitivity weight function $W_S$; low frequency gain $l_T = 0.2$, the high frequency gain $h_T = 1.8$, and the frequency of transition to significant model uncertainty $\omega_T = 120$ rad/sec for complementary sensitivity weight function $W_T$. The weights used are given by:

$$\left(W_S(s)\right)^{-1} = h_s \frac{s+\omega_s l_s}{s+\omega_s h_s} \quad (45)$$

$$W_T(s) = h_T \frac{s+\omega_T l_T}{s+\omega_T h_T} \quad (46)$$

Fig. 6 shows the $k_d - k_p$ solution region obtained for this multi-objective design and $(k_d, k_p)$ are selected as (0.07, 0.2). It can be seen from Fig. 7 and Fig. 8 that the corresponding frequency responses satisfy the phase margin constraint, and the mixed sensitivity constraint is also satisfied with the chosen controller parameters as the magnitude plot is below the 0 dB $((|W_S S| + |W_T T|) = 1)$ line.

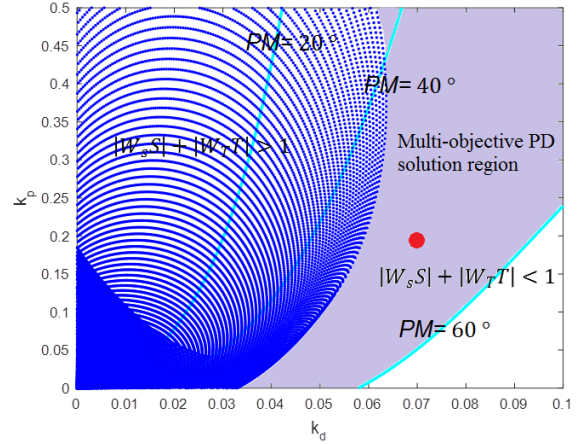

Fig. 6. Multi-objective discrete time PD controller design

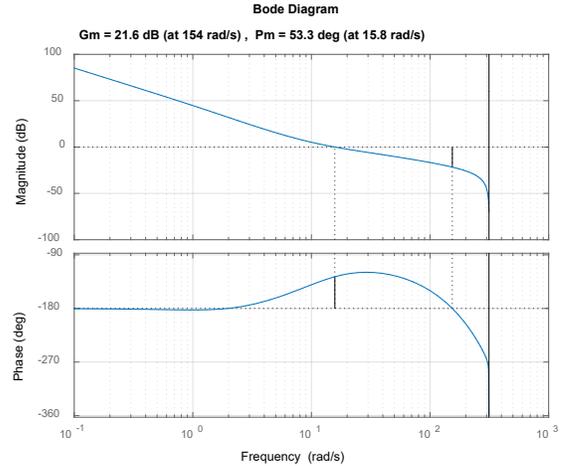

Fig. 7. Phase margin constraint

The designed digital PD feedback controller where $k_d = 0.07$, $k_p = 0.2$ and T =0.01 sec is used for autonomous vehicle path following. The simulation results of vehicle trajectory and vehicle lateral deviation error with the designed digital PD controller are displayed in Fig. 9 and Fig. 10, respectively. The desired path is an elliptical route and the total distance of one loop is 4,285 m. The vehicle drives at 60 km/h. It can be seen from Fig. 10 that the vehicle successfully tracks the desired path shown in Fig. 9 while there are small path following errors at corners.

## V. CONCLUSION

This paper introduced a parameter space approach based multi-objective digital PID controller design method. Absolute stability, phase margin constraint, gain margin constraint and mixed sensitivity constraint were treated. In an illustrative example, multi-objective robust digital PD controller gains were designed considering phase margin constraint and mixed sensitivity constraint simultaneously and used in the feedback control system of autonomous vehicle path following.

Simulation results show the effectiveness of the proposed digital controller design method. The approach used in this paper can be extended to design robust parameter space based disturbance observer control [3], [15]. The multi-objective digital PID controller design method of this paper can also be applied to different applications in future work [16-27].

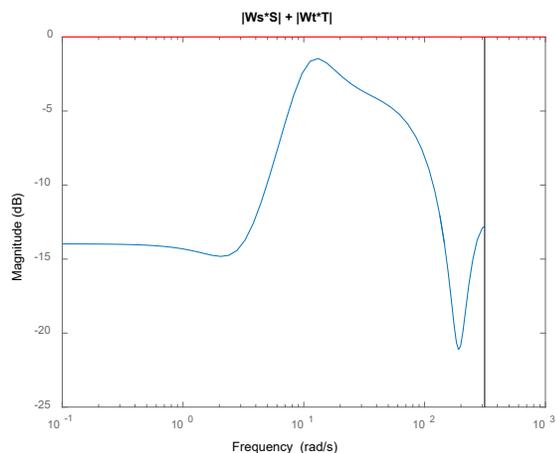

Fig. 8. Robust performance plot

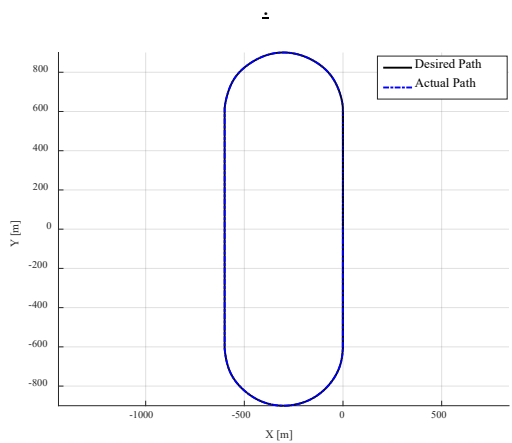

Fig. 9. Vehicle Trajectory

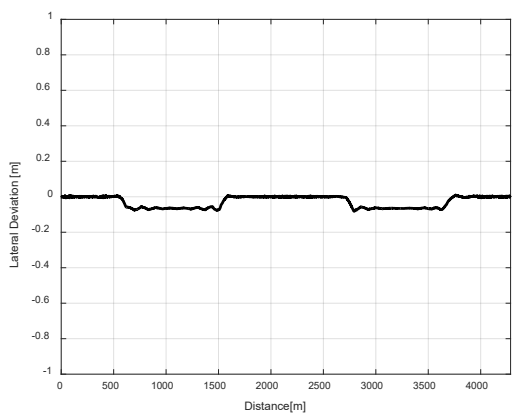

Fig. 10. Lateral Deviation

Wang and Guvenc: Multi-Objective Digital PID Controller Design in Parameter Space 7Programming Method," *IEEE Systems, Man and Cybernetics Conference*, İstanbul, October 10-13, pp. 4318-4324, 2010.
[18] Aksun-Guvenc, B., Guvenc, L., 2002, "The Limited Integrator Model Regulator and its Use in Vehicle Steering Control," Turkish Journal of Engineering and Environmental Sciences, pp. 473-482.
[19] Aksun-Guvenc, B., Kural, E., 2010, "Model Predictive Adaptive Cruise Control, IEEE International Conference on Systems, Man and Cybernetics, pp. 1455-1461.
[20] Aksun Guvenc, B., Guvenc, L., Ozturk, E.S., Yigit, T., 2003, "Model Regulator Based Individual Wheel Braking Control," IEEE Conference on Control Applications, İstanbul, June 23-25.
[21] Emekli, ME, Aksun Guvenc, B., 2016, "Explicit MIMO model predictive boost pressure control of a two-stage turbocharged diesel engine," IEEE transactions on control systems technology 25 (2), 521-534.
[22] Emirler, M.T., Kahraman, K., Sentrrk, M., Aksun Guvenc, B., Guvenc, L., Efendioglu, B., 2013, "Estimation of Vehicle Yaw Rate Using a Virtual Sensor," International Journal of Vehicular Technology, Vol. 2013, ArticleID 582691.
[23] Emirler, M.T., Guvenc, L., Aksun-Guvenc, B., 2018, "Design and Evaluation of Robust Cooperative Adaptive Cruise Control Systems in Parameter Space," International Journal of Automotive Technology, Vol. 19, Issue 2, pp. 359-367.
[24] Aksun-Guvenc, B., Guvenc, L., 2002, "Robust Steer-by-wire Control based on the Model Regulator," Joint IEEE Conference on Control Applications and IEEE Conference on Computer Aided Control Systems Design, Glasgow, pp. 435-440.
[25] Guvenc, L., Srinivasan, K., 1994, "Friction Compensation and Evaluation for a Force Control Application," Journal of Mechanical Systems and Signal Processing, Vol. 8, No. 6, pp. 623-638.
[26] Emirler, M.T., Wang, H., Aksun-Guvenc, B., Guvenc, L., 2015, "Automated Robust Path Following Control based on Calculation of Lateral Deviation and Yaw Angle Error," ASME Dynamic Systems and Control Conference, DSC 2015, October 28-30, Columbus, Ohio, U.S.
[27] Ozcan, D., Sonmez, U., Guvenc, L., 2013, "Optimisation of the Nonlinear Suspension Characteristics of a Light Commercial Vehicle," International Journal of Vehicular Technology, Vol. 2013, ArticleID 562424.